\documentstyle[12pt]{article}
\input{epsf}
\def\beq{\begin{equation}}
\def\eeq{\end{equation}}
\def\bea{\begin{eqnarray}}
\def\eea{\end{eqnarray}}
\def\bq{\begin{quote}}
\def\eq{\end{quote}}

\def\gappeq{\mathrel{\rlap {\raise.5ex\hbox{$>$}}
{\lower.5ex\hbox{$\sim$}}}}

\def\lappeq{\mathrel{\rlap{\raise.5ex\hbox{$<$}}
{\lower.5ex\hbox{$\sim$}}}}

\date{}
 
\begin{document}
 
\setcounter{page}{0}
\thispagestyle{empty}
 
\begin{flushright}
CERN-TH/97-46\\
hep-ph/970????\\
\end{flushright}
\vspace{0.2cm}
 
\begin{center}
{\Large \bf Signatures of Parton Exogamy in}\\
{\Large \bf $e^+ e^- \rightarrow W^+ W^- \rightarrow \, hadrons$}\\
\end{center}
\bigskip
\smallskip
 
\begin{center}
 
{\large
{\bf John Ellis{\boldmath $^a$} and Klaus Geiger{\boldmath $^b$}}}
\bigskip
 
$^a${\it Theoretical Physics Division, CERN,
CH-1211 Geneva 23, Switzerland:\\ 
John.Ellis@cern.ch}
 
$^b${\it Physics Department, BNL, Upton, N.Y. 11973, U.S.A.:\\ 
klaus@bnl.gov}
\end{center}
\vspace{0.5cm}
 
\begin{center}
{\large {\bf Abstract}}
\end{center}
\smallskip
We propose possible signatures of `exogamous' combinations between 
partons in the different $W^+$ and $W^-$ hadron
showers in $e^+e^- \rightarrow W^+W^-$ events with purely
hadronic final states. Within the space-time model
for hadronic shower development that we have proposed previously, we find
a possible difference of
about 10 \% between the mean hadronic multiplicity in such purely
hadronic final states and twice the hadronic multiplicity in events
in which one $W^{\pm}$ decays hadronically and the other leptonically,
i.e., $<N_{had} (2W)> \ne 2 <N_{had}(W)>$,
associated with the formation of hadronic clusters by `exogamous' pairs
of partons. We discuss the dependence of this possible difference in
multiplicity on the center-of-mass energy, on
the hadron momenta, and on the angular separation
between the $W^{\pm}$ dijets. If it
were observed, any such multiplicity difference would 
indicate that the $W^{\pm}$ do not hadronize independently,
and hence raise questions
about the accuracy with which the $W^{\pm}$ mass could be determined
from purely hadronic final states.

\noindent
 
\vspace{0.5cm}
 

\newpage
 

One of the central aspects of the experimental programme at LEP 2
\cite{LEP2},
is the study of
the reaction $e^+e^- \rightarrow W^+W^-$. This will permit more
detailed investigation of $W^{\pm}$ production and decay than has been
possible previously, enabling more precise measurements of the
three-gauge-boson couplings and the $W^{\pm}$ mass $M_W$. There
are three main ways in which the latter can be measured at LEP 2:
a) using the value of the $W^+W^-$ cross section at an energy close to
threshold, b) kinematic fits to higher-energy events in which one
$W^{\pm}$ decays leptonically and the other hadronically, and
c) kinematic fits to events in which both $W^{\pm}$ decay hadronically.

The last of these is promising because of the high statistics that
high-energy LEP 2 running is expected to produce, and because of
the relative sophistication and accuracy of programs for analyzing
events with four hadronic jets. However, concern has been
expressed in the literature that these hadronic jet-mass measurements
might be vulnerable to shifts in the apparent mass of the
$W^{\pm}$ due to physical interference between the hadronic decay
products that emerge from their two initial dijet systems:
$W^+ \rightarrow q_1 \bar{q}_2$ and $W^- \rightarrow q_3 \bar{q}_4$.
Since
the two initial colour-singlet systems
$q_1 \bar{q}_2$ and $q_3\bar{q}_4$ 
are produced essentially on top of each other at LEP~2, and 
evolve almost simultanously, it is natural
to be concerned
that the quarks and gluons from the two sources may
cross-talk, thereby altering the naive picture of independent
evolution and fragmentation. Indeed, comparison between
purely hadronic decays $W^+W^-\rightarrow q_1\bar{q}_2 + q_3\bar{q}_4$
and semileptonic decays
$W^+W^-\rightarrow q_1\bar{q}_2 + \ell \nu$ with the same kinematics
could be a sensitive probe of the confinement dynamics.
However, the current consensus is that
large interference effects are unlikely to
be generated during the perturbative phase \cite{khoze92,khoze93,khoze94}
of parton shower
development that follows immediately on the electroweak decays of the
$W^{\pm}$, but the question remains open whether significant
interference effects might appear during the subsequent non-perturbative
hadronization phase \cite{GPZ,SK,GH,BW,LL,ms40}. 
These might result either from effects on the
parton-to-hadron conversion process - due to the fact that the decay
products of the $W^{\pm}$ overlap in space and time,  hadronic clusters
might be formed by coalescence of partons from the $W^+$ and $W^-$
showers, a possibility we term {\it exogamy} - or from the
statistics of final-state hadrons - most importantly Bose-Einstein
correlations among pions.

In a previous paper \cite{ms40}
we have analyzed possible exogamy effects in the parton-to-hadron
conversion process using a model for the space-time development of
hadronic showers which is based on perturbative QCD transport theory
\cite{ms3942}
for the evolution of partons, followed by an ansatz for
parton-to-hadron conversion \cite{ms37}
that is based on a spatial criterion for
confinement. 
The latter ansatz rests on the insight that 
hadronization of parton showers in hard QCD processes appears
to be a `local' phenomenon \cite{LPHD}, in the sense that it is
determined by the favour and colour degrees of partons 
which are close-by in phase space. Such nearest-neighbour
partons  most preferably tend to form colour-singlet
pre-hadronic clusters \cite{preconf,marchesi}
out of which final-state hadrons emerge.
A key implication is that details of the space-time
structure of the evolving parton ensemble, given by its time-dependent
phase-space density,  should be reflected
in the final hadronic yield; however, the details of the
actual parton-hadron conversion mechanism in a given phase-space cell
should be local and universal in nature. Consistently with this picture,
we incorporated no {\it a priori}
prejudice that hadronization should prefer `endogamous' unions of
partons from the same $W$ decay shower, and we found a large
fraction of `exogamous' unions between partons from different
$W$ showers. The identities of individual $W^{\pm}$ decay
products are therefore not well defined. Moreover, whereas
previous analyses of the possible implications
for $M_W$ of colour recombination effects \cite{GPZ,SK,GH,BW,LL} 
had suggested uncertainties of 
less than 100 MeV, using standard jet algorithms we found 
\cite{ms40} shifts in $M_W$
of several hundred MeV
between our favoured model for parton-to-hadron conversion and other
scenarios based on our framework for the space-time development of
hadronic showers, as well as a hypothetical scenario in which the $W^{\pm}$
decays are widely separated in space.
Our hypothesis that partons are unbiased in the selection of their future
hadronization partners is in 
contrast to the standard string
picture, in which the colour charges of the initial quark $q_1$ and antiquark 
${\bar q}_2$  predetermine 
at the space-time point of production the colour flow all the way into the remote
future when finally hadrons are formed.
The provocative question we raise is:
{\it
why, or to what extent, should  the initial quark $q_1$ remember at a time
several fm/c later that its original colour partner was
$\bar{q}_2$, and that it is supposed to form a string, or hadron chain,
with the latter, when in the meantime plenty of other quark and
gluon colour charges may have been
produced in the region between the receding $q_1$ and 
${\bar q}_2$?
}
\footnote{The loss of colour memory has been 
studied quantitatively for the extreme case
of a high-density QCD plasma, in which case it was found that
the colour relaxation time during
which the parton colour charges are completly randomized
was extremely short:
$\Delta t  \, \lower3pt\hbox{$\buildrel <\over\sim$}\, 0.1$ 
$fm$~\cite{selikov93}.}
\smallskip

In view of the potential significance for the LEP 2 experimental
programme of such an uncertainty, in this paper we explore in more
detail some implications of our space-time picture of hadronization,
looking in particular for signatures that might provide
`early warning' of possible large interference effects due to such
exogamous parton combinations. Specifically, we observe that
these may cause a significant difference between the mean
multiplicity in a single $W^{\pm}$ hadronic decay and half the mean
multiplicity
in events with a $W^+W^-$ pair each of which decays hadronically, i.e.,
$<N_{had}(2W)> \ne 2 <N_{had}(W)>$. We also explore the possible
distribution of any such multiplicity difference in the final-state
hadronic phase space, finding that it is particularly enhanced 
for small hadron momenta, and that the effect has a strong dependence
on the relative angles of the dijets from the $W^+$ and $W^-$. The
difference in the total mean multiplicity could
be as large as 10 \%, which might be detectable with 
relatively small event  samples.
The effect that we find decreases slowly with the
increasing center-of-mass energy, but is still several \% even at the
highest possible LEP 2 energies.

Referring to \cite{ms40} for details, we 
summarize briefly here the essential concepts of our 
space-time model for parton shower development and hadronization \cite{ms44}:

The {\it parton shower dynamics} is described by 
conventional perturbative QCD evolution Monte
Carlo methods, with the added feature that we keep track of the spatial
development in a series of small time increments. Our procedure
implements perturbative QCD transport theory in a manner consistent 
with the appropriate quantum-mechanical uncertainty principle,
incorporating parton splitting and recombination~\footnote{The latter
is not significant during the perturbative phase of the hadron
shower in $e^+e^-$ annihilation.}. 
In the rest
frame of each $W^{\pm}$, each off-shell parton $i$ in the shower
propagates for a time $\Delta t_i$ given in the mean by $<\Delta t_i> =
\gamma_i \tau_i = E_i /k^2_i = x_i M_W /2
k^2_i$, where $k^2_i$ is the parton's squared-momentum virtuality,
and $x_i = E_i / M_W$ its longitudinal energy fraction, during which it
travels a distance $\Delta r_i = \Delta t_i \beta_i$. The $n$'th step in
the
shower cascade is completed after a time $t_n = \Sigma^n_{i=1} \Delta
t_i = \frac{1}{2}M_W \Sigma^n_{i=1} x_i / k^2_i$. 
This means that the typical
time after which a parton with momentum fraction $x$ reaches a low
virtuality $k_0^2 = {\cal O}(\Lambda_{QCD}^2)$ is $<t(x, M_W^2)> \sim
(x M_W / k^2_0)\, \hbox{exp}(-b \sqrt{\hbox{ln}x})$. Thus, soft partons
are expected to hadronize first, in a conventional inside-outside cascade
\cite{marchesi}.

The {\it parton-hadron conversion}, on the other hand,  is
handled using a {\em strictly spatial} criterion for confinement, with
a simple field-theoretical ansatz used to estimate the probability $P(R)$
that a spatial region of given size $R$ will make a transition 
from the parton phase to
the hadron phase. At each time step in the shower development, every
parton that is further from its neighbours than a certain critical
distance $R_c$ estimated using a simple field-theoretic model for
parton-hadron conversion~\cite{CEO}, is
assigned the corresponding probability $P(R)$, estimated within the same
model, to combine with its
nearest-neighbour parton to form a hadronic cluster, possibly
accompanied (in our favoured `colour-full' hadronization scenario) 
by one or more partons to take correct account of the colour
flow. The resulting hadronic clusters are then allowed to decay into
stable hadrons according to the particle data tables.

In the application to $e^+e^- \rightarrow W^+W^-$ that we pursue here,
it is important to stress again that at no moment in this shower
development
do we make any distinction between the decay products of the $W^+$ and
$W^-$. In any given $e^+e^-$ annihilation event, an exogamous
pair of partons from different $W^{\pm}$ decays have the same
probability of conversion to hadrons as 
do an endogamous pair of partons from the same
$W^{\pm}$ decay, at the same spatial separation. This philosophy of
parton-hadron conversion may be distinguished from the mainstream
approach in which each $W^{\pm}$ decay is assumed {\it a priori}
to form a string, the criterion for cluster formation is formulated
essentially in momentum space, and there are a limited number of
cross connections between the different strings~\cite{LEP2}.
This difference in philosophy is not of academic interest, 
since, for $e^+e^- \rightarrow W^+W^-$ production at LEP 2, the
partonic showers of $W^{\pm}$ decays are superposed. Before decay,
each $W^{\pm}$ travels a distance $r^{\pm}$ given in the mean by
$r^{\pm} = \gamma^{\pm} \tau^{\pm} \beta^{\pm}$, where
$\tau^{\pm} = 1 / \Gamma_W$. Numerically, this distance is very short
at LEP 2, namely, of order 0.05~fm (0.1~fm) at a center-of-mass energy 
$E_{CM} = 170\;(200)$ GeV, and the boost of the $W^{\pm}$ at higher
energies will not be sufficient to separate their decays by more than
1~fm at least until $E_{CM} > 1$ TeV.
Although
the leading high-momentum partons separate rapidly, in general, 
because the directions of the 
$W^{\pm} \rightarrow q_i{\bar q}_j$ decay axes are different, 
the two clouds of low-momentum partons
can be expected to overlap, leading to a substantial number of
exogamous unions, as seen clearly in Fig.~18 of Ref. \cite{ms40}. These may
also occur among higher-momentum partons,
if the angles between the $W^{\pm}$ dijets 
decay axes are sufficiently small, as we discuss below.

In our previous paper \cite{ms40}, we analyzed the potential significance of
such exogamous unions for the experimental determination of $M_W$
from purely hadronic final states. We applied standard jet-finding
and dijet mass-estimation algorithms to hadronic final states
obtained using three different variants of our space-time model for
parton-to-hadron conversion, 
finding mass shifts  $\delta M_W$ of up
to several hundred MeV, compared to independent $W^{\pm}$ decays.
For example, in our  `colour-full' scenario in which  
partons in any colour combination can coalesce to colour-singlet clusters 
through additional non-perturbative gluon emission, 
we found a shift of 0.27~GeV in $\langle M_W \rangle$ between realistic
overlapping $W^+W^-$ decays as compared to
hypothetical independent $W^+W^-$ decays.

In the following we will 
focus on different kinematical signatures of cross-talk
between the $W^{\pm}$ decay products, in particular the multiplicity
dependence and the shape of particle spectra.
The results that we will discuss were obtained from simulations with
the Monte Carlo implementation \cite{ms40,ms44} of our model.
For the full range of LEP~2 energies, $E_{cm} = 162 -200$ GeV,
we have studied two distinct situations: 
first, the physically {\it realistic}
overlapping evolution of cross-talking parton showers
as they occur in events where both $W$'s decay hadronically
via $W^+W^-\rightarrow q_1\bar{q}_2 + q_3\bar{q}_4$,
and, secondly, 
the non-interfering
{\it independent} evolution of superposed dijets from decays of $W^+$ and
$W^-$,
mimicking events where one of the $W$'s decays hadronically and 
the other one semi-leptonically. Technically, the latter case is
implemented by separating physically the two $W^{\pm}$ decay vertices
by a large distance $>> 1$ fm, and should resemble two
superposed $W^+W^-\rightarrow q_1\bar{q}_2 + \ell\nu$ decays.
In the following we use the terms 
$1\times 2 W$ for {\it purely hadronic} decays and $2 \times 1 W$ for
{\it superposed semileptonic} decays.

Fig.~1 displays potential gross signatures of cross-talk between the
$W^\pm$
hadronic decay showers in the energy range $E_{cm} = $162 - 200 GeV:
the top part exhibits a significant difference between 
the mean hadronic
multiplicities $<N_{ch}>$ in {\it realistic} overlapping and {\it 
independent} separated $W^\pm$ decays
of about 10 \%, which persists with very weak $E_{cm}$-dependence
throughout the entire LEP~2 energy range.
The hadronic decays of both $W^+$ and $W^-$ (the $1 \times 2 W$ case) 
yield generally about 10\% lower charged multiplicity than 
twice the case when one $W^{\pm}$ decays semi-leptonically (the $2 \times
1 W$ case).
The slight increase with $E_{cm}$ of $\langle N_{ch} \rangle$ 
originates from a weak growth of $\langle M_W\rangle$ which
sets the initial evolution scale for the parton showers. This is
due to the gradual relaxation of the kinematical constraint that
favours the low-mass side of the $W^{\pm}$ Breit-Wigner close to the
nominal $W^+W^-$ threshold. The bottom part of Fig.~1 displays the
difference in the mean transverse momenta $<p_{T~ch}>$ 
of charged particles, measured with respect to the thrust axis
\footnote{Thrust is defined as usual,
$T = \max_{|\vec n|} \sum_i \vec{n}\cdot\vec{p}_i/\sum_i |\vec{p}_i|$
with respect to all particles $i$ in an event (not within the 2 dijets individually),
and the thrust axis is given by the unit vector $\vec{n}$ for which the maximum 
is attained.},
between the realistic $1 \times 2 W$ and independent $2 \times 1 W$ cases.
The purely hadronic events give larger mean transverse momenta to the
final-state particles than do events with one semi-leptonic decay,
mainly because the total transverse momentum 
is of similar magnitude in each case, but is distributed among fewer
particles in the lower-multiplicity $1 \times 2 W$ case.

\begin{figure}
\epsfxsize=350pt
\centerline{ \epsfbox{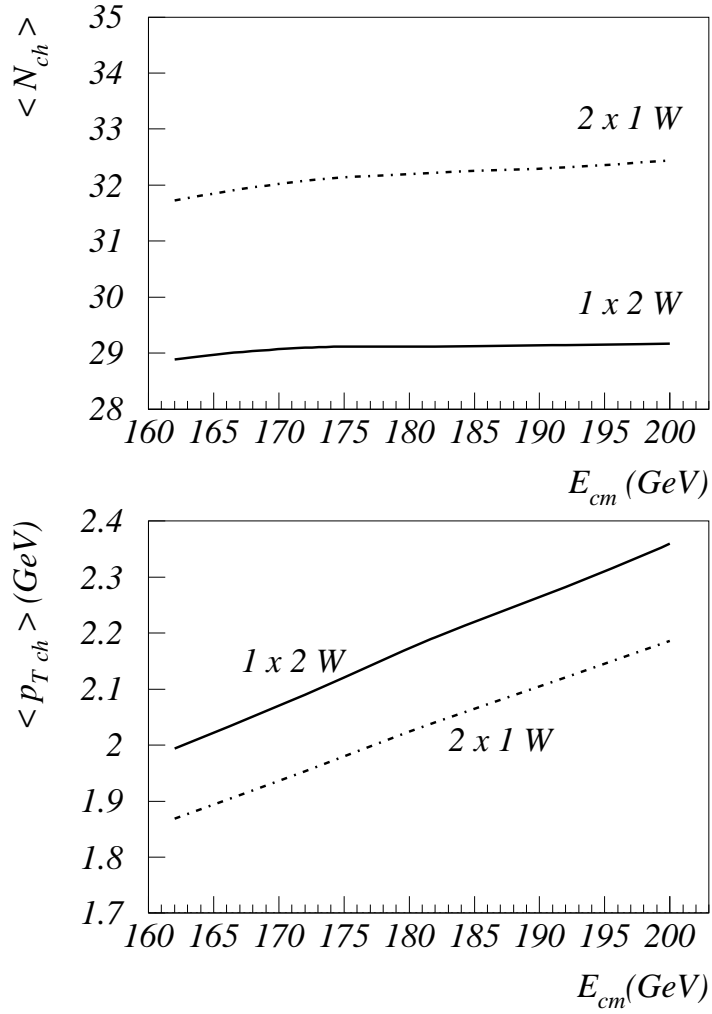} }
\vspace{-1.5cm}
\caption{
Energy-dependences ($E_{cm}=s^{1/2}$) of the average charged
multiplicity and the mean transverse momentum per charged particle in
realistic cross-talking $W^+W^-$ decay events ($1\times 2 W$),
compared with independent $W^{\pm}$ decay events ($2 \times 1 W$).
\label{fig:fig1}
}
\end{figure}

We see that these possible
differences persist throughout the LEP 2 energy range between 162 and 200
GeV, consistent with our earlier suggestion that these effects may not
disappear before $E_{cm} \sim$ TeV. The plots in Fig. 1 (and in
the following figures) are for our preferred
`colour-full' hadronization scenario~\cite{ms40}, in which we take
account of both the spatial separation and the colour matching of
of partons coalescing to form a pre-hadronic colour-singlet
cluster. For comparison, we have also studied the naive `colour-blind'
scenario,
which is based solely on the spatial nearest-neighbour criterion
for parton coalescence, irrespective of the colour degrees of freedom.
The two prescriptions, `colour-full' and `colour-blind', are confronted
in Table 1, where we see that some difference between the $1 \times 2 W$
and $2 \times 1 W$ values of
$<N_{ch}>$ is always present at the several \% level, despite detailed
differences between the hadronization scenarios.
\begin{figure}
{\small
\begin{center} 
\begin{tabular}{|c|cc|cc|cc|}
\hline
\hline
  mode  
&
 \multicolumn{2}{c|}{$\;\;\;\;\;\;$ $\langle N_{ch} \rangle$ ,  all $\;\;\;\;\;\;\;\;$}
&
 \multicolumn{2}{c|}{$\;\;\;\;\;\;$ $\langle N_{ch} \rangle$ , $ | \vec{p} | \le 1$ GeV $\;\;\;\;\;\;\;\;$}
&
 \multicolumn{2}{c|}{$\;\;\;\;\;\;$ $\langle N_{ch} \rangle$ , $ | \vec{p} | \le 0.5$ GeV $\;\;\;\;\;\;\;\;$}
\\
               & $\;\;\;\;\;$     tot   &   $|y|\le 1$ 
               & $\;\;\;\;\;$   tot    &  $|y|\le 1$  
               & $\;\;\;\;\;$   tot   &   $|y|\le 1$  \\ 
\hline
\hline
             & & & & & & \\
         a)  & $\;\;\;\;\;$     31.6  &  24.8 &  $\;\;\;\;\;$ 13.7 &   11.9  & $\;\;\;\;\;$  7.5 &    6.6 \\
         b)  & $\;\;\;\;\;$     28.9  &  22.7 &  $\;\;\;\;\;$ 12.4 &   10.7  & $\;\;\;\;\;$  7.0 &    6.1 \\
         c)  & $\;\;\;\;\;$     26.8  &  21.3 &  $\;\;\;\;\;$ 10.2 &    8.6  & $\;\;\;\;\;$  4.9 &    4.3 \\
         d)  & $\;\;\;\;\;$     24.7  &  19.8 &  $\;\;\;\;\;$  9.3 &    7.8  & $\;\;\;\;\;$  4.6 &    4.0 \\
             & & & & & & \\
\hline
\hline
\end{tabular}
\end{center}

\begin{center}
\begin{tabular}{|c|cccc|}
\hline
\hline
         mode    &       $\langle p_{\perp\,ch} \rangle$ 
                 &   $\langle p_{\perp\,ch}^2 \rangle$  
                 & $\langle E_{ch} \rangle$  
                 &  $\langle p_{z\,ch} \rangle$ \\
                 &  $\;\;\;\;\;\;\;$     (GeV)  $\;\;\;\;\;$
                 &  $\;\;\;\;\;\;\;$  (GeV$^2$) $\;\;\;\;\;$
                 &  $\;\;\;\;\;\;\;$     (GeV)  $\;\;\;\;\;$
                 &  $\;\;\;\;\;\;\;$     (GeV)  $\;\;\;\;\;$ \\
\hline
\hline
                 & & & & \\
         a)      &         1.89  &      12.9  &      2.51 &     1.25 \\
         b)      &         1.96  &      13.9  &      2.58 &     1.26 \\
         c)      &         2.26  &      17.3  &      3.01 &     1.52 \\
         d)      &         2.40  &      19.5  &      3.15 &     1.57 \\
                 & & & & \\
\hline
\hline
\end{tabular}
\end{center}
}
\nopagebreak \noindent
{\bf Table 1:}
Summary of some simulation results at $E_{cm} = 162$ GeV.
The numbers for particle multiplicities and momenta refer to
the following cases:
a)  $2 \times 1 W$  -  colour-full,  
b)  $1 \times 2 W$  -  colour-full,  
c)  $2 \times 1 W$  -  colour-blind,
d)  $1 \times 2 W$  -  colour-blind.
Here, `colour-full' is our preferred hadronization scenario, which
accounts for both the spatial separation and the colour matching of
partons coalescing to form a pre-hadronic colour-singlet
cluster, whereas `colour-blind' is a naive hadronization scenario,
based solely on the spatial nearest-neighbour criterion
for parton coalescence, irrespective of the colour degrees of freedom:
see~\cite{ms40} for details. Since we have not attempted to fine-tune
either of our colour-full and clour-blind hadronization scenarios,
the absolute values of the multiplicities shown are not to be taken
as precise predictions: the significance lies in the differences between
the $2 \times 1 W$ and $1 \times 2 W$ cases for the two scenarios,
{\it i.e.}, the differences a) - b) and c) - d).
\end{figure}

Since it is not evident that a 10\% effect in the mean charged
multiplicity can be measured with limited statistics, we have
analyzed, within our model, the statistical fluctuations to be
expected in small samples of $W^+W^-$ events.
Specifically, we find that the  charged particle multiplicity
averaged over event samples of 100 events each,
shows a strongly peaked distribution around $\langle N_{ch} \rangle$
averaged over all event samples, with a small width
corresponding to fluctuations of
$\delta <N_{ch}>  \, \lower3pt\hbox{$\buildrel <\over\sim$}\,\pm 2$.
This indicates that the effects in Fig.~1 are only marginal with the
present statistics from $E_{CM} = 161, 172$ GeV, but should be
resolvable with the experimental data obtained during 1997.
\smallskip

As originally suggested in Ref. \cite{GPZ}, and in line with the
previous qualitative arguments, one would expect the bulk of any
difference in charged multiplicity to appear in kinematical
configurations where exogamy is most prevalent, namely
(i) when the two dijets from the $W^-$ and $W^+$ decays are
produced in an `anti-collinear' configuration with the initial
$q_1$ and ${\bar q}_4$ emerging
with a small relative angle (and similarly for
${\bar q}_2$ and $q_3$), and (ii) in the central
rapidity region of small particle rapidities (momenta), where most
of the gluonic off-spring is produced. In order to investigate these
expectations, we measure momenta 
with respect to the axis $\widetilde{z}$ that cuts the two dijet axes in
half.
The axis $\widetilde{z}$ is constructed,
as illustrated in Fig. 2, from the the
knowledge of the initial two jet axes
of $q_1\bar{q}_2$ and $q_3\bar{q}_4$ (not from the final hadronic jets),
where the relative angle $theta$ is defined as the angle between 
the quark of one dijet and the antiquark of the other dijet, i.e.,
between $q_1$ and $\bar{q}_4$, or equivalently $\bar{q}_2$ and $q_3$.
\smallskip

Fig. 3 shows that the mean charged multiplicity depends
strongly on the relative
angle $\theta$ between the cross-talking dijets in the $1 \times 2 W$
case, shown by the solid line. 
For comparison, the dashed line is for
the $2 \times 1 W$ case of independent dijet evolution. This is flat to
within our statistical errors, indicating
that there is no angular dependence in this case, as expected. The wiggles
in this line serve to emphasize the significance of the variation seen in
the cross-talking $1 \times 2 W$ case. Experimentally, it is not feasible
to separate efficiently the kinematical configuration of anti-collinear
dijets from that of collinear dijets, in which $q_{1,3}$ emerge in similar
directions, as do ${\bar q}_{2,4}$. Therefore the effect seen in Fig. 3
will be diluted experimentally by folding together the configurations
$\theta$ and
$180^o - \theta$.

\begin{figure}
\epsfxsize=400pt
\centerline{ \epsfbox{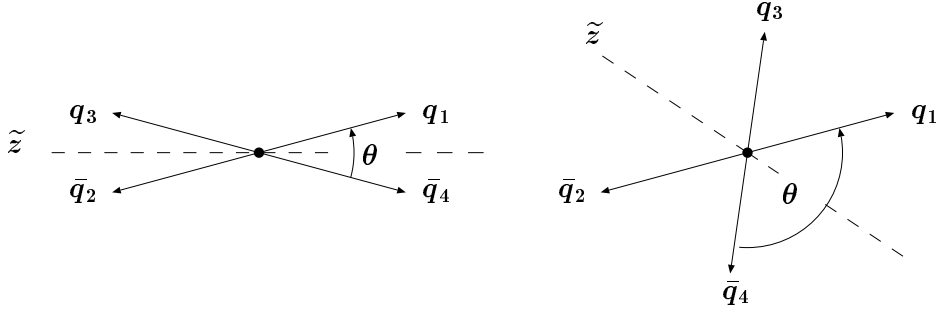} }
\vspace{-11.0cm}
\caption{Illustration of the definitions of the axis ${\widetilde z}$
and the dijet separation angle $\theta$ between $q_1$ and ${\bar q_4}$
(equivalently, between ${\bar q}_2$ and $q_3$) from the decays
$W^+ \rightarrow q_1 {\bar q}_2$ and $W^- \rightarrow q_3 {\bar q}_4$.
Small- and large-angle configurations are shown.
\label{fig:fig2}
}
\end{figure}

\begin{figure}
\vspace{-6.0cm}
\epsfxsize=400pt
\centerline{ \epsfbox{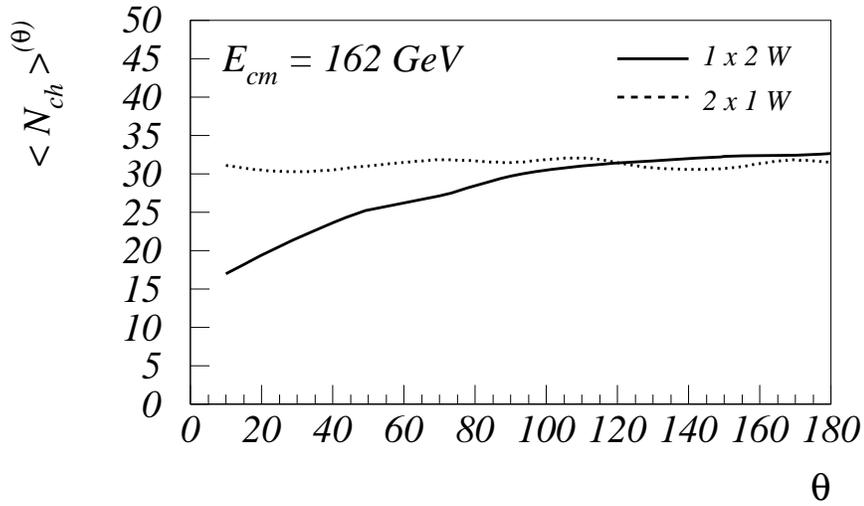} }
\vspace{-6.0cm}
\caption{
Mean charged-particle multiplicity as a function of the 
relative angle $\theta$ between the two initial dijets defined
in Fig.~2. The solid
(dotted) curve refers to the cross-talking (independent)
evolution
of the $W^+$- and $W^-$-initiated parton showers.
\label{fig:fig5}
}
\end{figure}

We now look into more details of the hadron distributions in
$W^{\pm}$ final states. We see in the top panel of Fig. 4 that 
the full multiplicity distributions have similar shapes 
in the $1 \times 2 W$ and the $2 \times 1 W$ cases, though with
different mean values of
28.9 and 31.6, respectively, in our preferred colour-full
hadronization scenario. However, we also see in
Fig. 4 that most of the hadronic multiplicity
differences
between the cross-talking evolution ($1 \times 2 W$) and 
the independent evolution ($2 \times 1 W$) indeed occur
(i) at small relative angles 
$\theta   \, \lower3pt\hbox{$\buildrel <\over\sim$}\,30^o$ as
already seen in Fig. 3,
and (ii) for particles with $| y | \le 1$ (corresponding 
mostly to small momenta 
$|  \vec{p} |\, \lower3pt\hbox{$\buildrel <\over\sim$}\,1$ GeV).
Specifically, as seen in the top panel of Fig. 4, small-angle events with
$\theta < 30^o$ yield a shifted multiplicity distribution with the much
lower mean multiplicity of 21.5.
The rapidity spectra of charged particles shown in the bottom panel of
Fig. 4 diagnoses these differences,
showing that, in the average over all angular configurations, the
$1 \times 2 W$ and the $2 \times 1 W$ rapidity distributions have
characteristically different shapes.
The $2 \times 1 W$ case has the typical humped shape of ordinary $q\bar q$
jet events,
whereas the $1 \times 2 W$ case exhibits more of a plateau in the
central rapidity region, indicating a significant suppression of soft
gluon emission. This effect is particularly marked
in small-angle events, where the  depletion in the region $| y | \le 1$
is very prominent. Notwithstanding the suggestive spectra in Fig. 4, 
we emphasize that there are
significant fractional differences in the rapidity distributions even when 
$\vert y \vert > 1$, as also shown in Table~1. 
Indeed, although the fractional
difference in multiplicity may be enhanced in the central region, it does not
seem that this enhancement is large enough to improve noticeably the
measurability of such a difference in an experiment with limited statistics,
such as are available from the LEP runs at 162 and 172 GeV.
\smallskip

\begin{figure}
\epsfxsize=350pt
\centerline{ \epsfbox{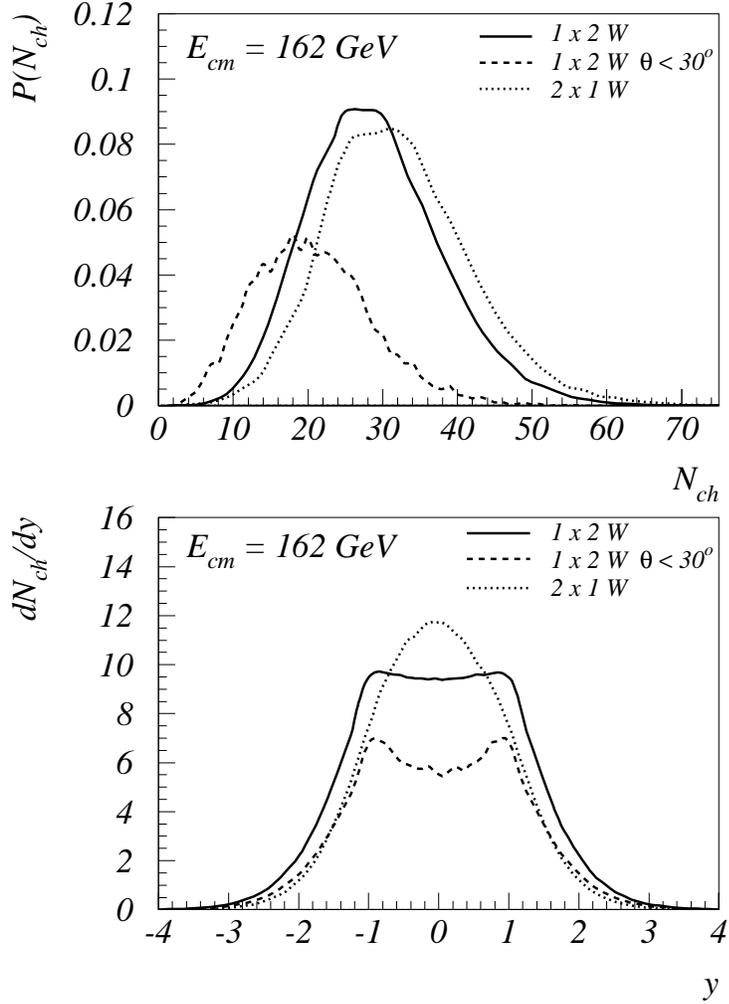} }
\vspace{-1.5cm}
\caption{
Charged-particle multiplicity distributions and spectra of rapidities
with respect to the $\widetilde{z}$ axis defined in Fig.~2. The
full (dotted) curves are for cross-talking (independent)
evolution of the $W^+$- and $W^-$-initiated parton showers. The
dashed lines show the corresponding distributions in small-angle events,
where exogamy effects are enhanced.
\label{fig:fig3}
}
\end{figure}

Why is the effect on the charged multiplicity a
{\it reduction}, rather than a {\it increase}?  
The results shown in Figs.~1, 3 and 4 could be explained by
the production either of lower-mass hadronic clusters, and/or of fewer
clusters, the effect becoming stronger for small-angle,
anti-collinear dijets, and for softer clusters.
In some loose sense, such an
effect may be thought of as reflecting increased ``efficiency" in the
hadronization process, possibly along the lines of the
basic quantum-mechanical principle of choosing the state of lowest
possible energy (or invariant mass). Perhaps the presence of two
cross-talking
dijets in the $1 \times 2 W$ case, with their spatially-overlapping
offspring, allows
the evolving particle system to reorganize itself more favorably in
the cluster-hadronization process, and to pick a state with smaller 
invariant mass than
in the $2 \times 1 W$ case corresponding to independent dijets
with no cross-talk. Indeed,
Fig. 5 shows that the mass spectrum of pre-hadronic
clusters from coalescing partons is in fact softer in the
$1 \times 2 W$ case, reflecting the fact that the availability of
more partons enables clusters to form from
configurations with lower invariant mass than
in the $2 \times  1 W$ case. Fig. 5 shows that this feature is also
enhanced at small relative
angles $\theta < 30^o$ of the two evolving dijets.
On the other hand, we have found that there is no significant
difference in the number of hadronic clusters produced
in the $1 \times 2 W$ and $2 \times 1 W$ cases~\footnote{For 
completeness, we report that in the $1\times 2 W$ case
at $E_{cm}= 162 \;(200)$ GeV an average  total of $N_{clus} = 24.8 \;(26.2)$
pre-hadronic clusters are produced, out of which
$N_{clus}^{(ex)} = 14.1 \;(16.8)$ are exogamous coalescence products.}.

\begin{figure}
\vspace{-6.0cm}
\epsfxsize=400pt
\centerline{ \epsfbox{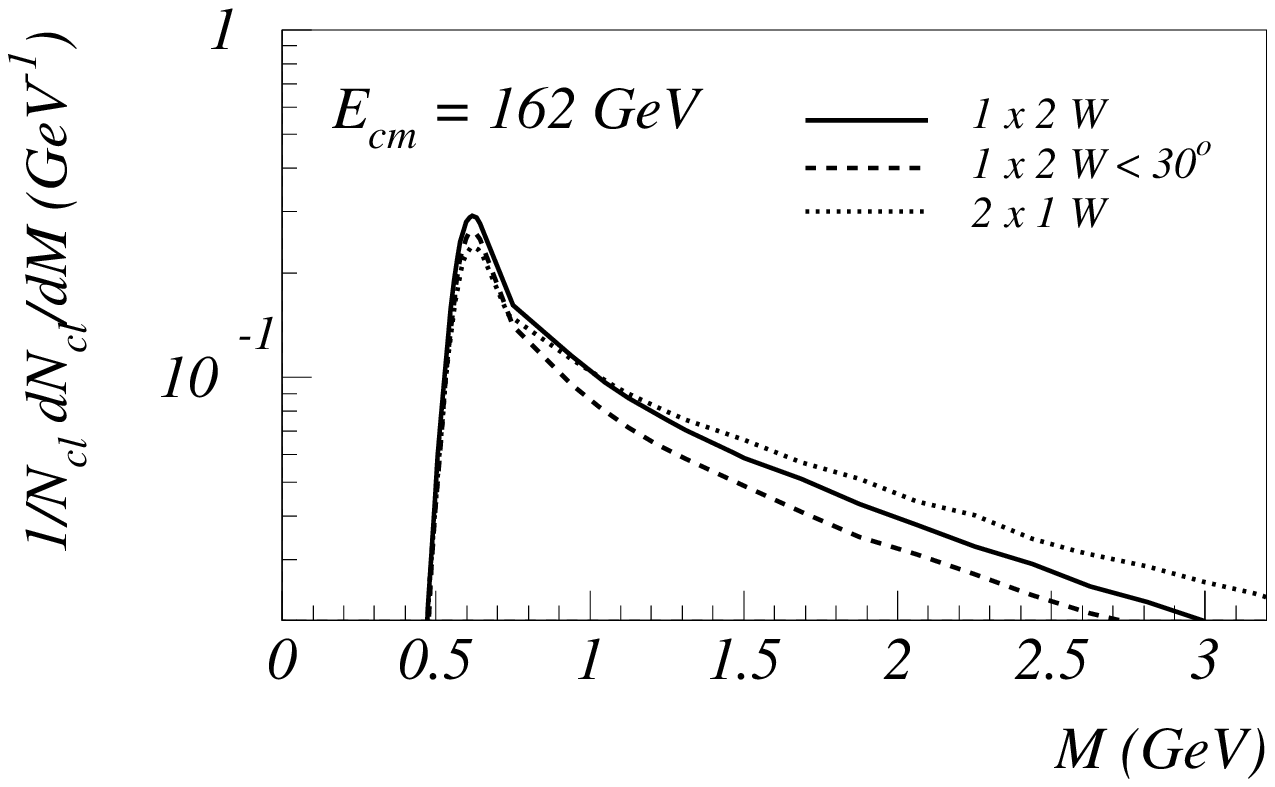} }
\vspace{-6.0cm}
\caption{
Mass spectra of pre-hadronic colour-singlet clusters formed from
coalescing partons during hadronization. The
solid (dotted) curves refer to cross-talking (independent)
evolution
of the $W^+$ and $W^-$ initiated parton showers.
The dashed lines shows the distribution for small-angle events,
where there is a more marked reduction in the fraction of high-mass
clusters. The mean values $\langle M \rangle$ are 
1.70 GeV for the small-angle case,
1.85 GeV for the cross-talking $1 \times 2 W$ case averaged over all
angles, and 2.2 GeV for the case of independent evolution.
\label{fig:fig4}
}
\end{figure}

\medskip

To render these differences between the overlapping
$1\times 2 W$ events and the independent $2\times 1 W$ events
more plausible, let us consider in more detail
an idealized configuration in which the second $W$ decays along an axis
parallel to that of the first $W$ decay, with $q_1$ 
accompanying ${\bar q}_4$, {\it i.e.}, $\theta = 0$ in Fig.~2.
In such a case, one would
expect strong interference between the comoving $\bar q$ and $q$ jets.
Indeed, if they were sufficiently close, the parton showers would have many
coherent features. Consider, for example, the case in which ${\bar q}_4$
and $q_1$ happen to form a colour singlet: the hadronic
final state should then tend to consist of two small groups of
high-momentum
hadrons at each end of the common jet axis, with a large intermediate
rapidity gap. Conversely, if the comoving ${\bar q}_4$ and $q_1$ form a
colour
octet, there could be an octet colour ``string" between the two ends of the
rapidity distribution along the common jet axis. Asymptotically,
this should yield a
higher rapidity plateau. However, it is known from studies of
gluon jets at accessible energies that any such enhancement is not as
large as the factor of $9/4$
that would be required to counterbalance the (idealized) rapidity-gap
events~\footnote{In reality,
the emission of additional gluons in the process of shower evolution 
complicates this simple-minded picture.
The emitted gluons, most of which carry small energy fractions
and populate the central rapidity region,
have - as discussed in Ref. \cite{ms40} - substantial
space-time overlap, and are, in our model, unbiased 
in their choices of coalescence partners, not caring which
$W$ decay they emerge from.
These low-momentum  gluons
populate mostly the spatial region of the
$W^+$ and $W^-$ decay vertices, particularly if $E_{cm}$ is close to 
the $W^+W^-$ threshold, and therefore increase the probability
for exogamous reorganization of colour flow between the initial dijets.
}
This argument points in the direction of an overall multiplicity reduction,
as seen in Figs.~1, which should become
more marked if the $W^\pm$ dijet axes are closer together, 
as seen in Fig.~5. Moreover, this simplified picture suggests
the appearance of a double hump in the rapidity distribution for
small-angle events, with a suppressed plateau, as seen in the bottom panel
of Fig.~3. This picture is also consistent with the enhanced number of
events with a very low number of charged particles, seen in the top panel
of Fig.~3. Confirmation of this picture could be provided by the
observation of a significant number of large-rapidity-gap events, which
are known to be very rare in $Z^0$ decays~\cite{rapgap}.
\bigskip

In conclusion, this analysis has shown that exogamy between the partons
from different $W^{\pm}$ hadronic showers
may have as an observable signature differences 
between the hadron distributions
in $(W^+\rightarrow  q_1\bar{q}_2)~(W^-\rightarrow q_3\bar{q}_4)$ 
events and independent $W \rightarrow q {\bar q}$ decays, as observable in 
$(W^\pm\rightarrow q_1\bar{q}_2)~(W^\mp\rightarrow  \ell\nu)$
events. These differences may become apparent even before any possible
difference in the $W^\pm$ masses extracted from these different classes of
events. Any observed difference should certainly put one on guard concerning
the interpretation of the $W^\pm$
mass extracted from purely hadronic final states in $e^+e^-\rightarrow
W^+W^-$, in the absence of any better understanding of the hadronization
process.

As a final general remark, we note  that $e^+e^-\rightarrow
W^+W^-$ provides a uniquely clean environment for probing the development of
hadronization in time and space. Other measurements, notably those in
$e^+e^-\rightarrow Z^0\rightarrow$ hadrons, only observe the final state at
distances $\vert\underline{r}\vert \gg$ 1 fm and times $t \gg$  1 fm/c. On
the other hand, as we have emphasized, the second $W$ decay occurs in the
heart of the ``hot spot" produced by the first $W^\pm \rightarrow \bar qq$
decay, and hadronization may provide a sensitive probe of the details of the
core of the parton-shower development. There are other unstable particles
which might
also provide tools for such studies, notably the $t$ quark. The process
$e^+e^-\rightarrow \bar tt$ may exhibit many of the features discussed here,
and this reaction is on the agenda of future $e^+e^-$ colliders. However,
there the main emphasis on measuring $m_t$ will be using total cross-section
measurements, so the implications of exogamy for mass reconstruction will be
less crucial than for $e^+e^-\rightarrow W^+W^-$. Currently, $m_t$ is being
extracted from jet measurements in $\bar pp\rightarrow \bar ttX$. The
estimated experimental error in $m_t$ is relatively large, but exogamous
hadronization effects might also be significant in this process, and merit
study.

At the moment, given the inevitable model-dependence of treatments of
hadronization, including that presented here, we think that the possible
effects of exogamy are currently an area where experiment must lead the way.
However, we have provided in this paper indications of some possible
signatures
that experiment might seek.  However, we should repeat one cautionary
remark: the numbers and particle 
distributions given in this paper are estimates
obtained within our model, that has not been tuned to fit
$e^+e^ \rightarrow W^+ W^-$ data, or even details of event-shape
variables in $Z^0 \rightarrow q {\bar q}$ decay. We do not
claim high accuracy for our estimates: rather, the intent of
our paper has to provide qualitative 
suggestions for possible interesting
physical effects, and to motivate further analysis by both theorists
and experimentalists. In particular,
we hope that forthcoming results from LEP~2 will soon
cast light on the issues raised in this paper.
\bigskip

This work was supported in part by the D.O.E. under contract no.
DE-AC02-76H00016.
\bigskip


\end{document}